\documentclass[twocolumn,prl,preprintnumbers,amsmath,amssymb,superscriptaddress]{revtex4}

\usepackage{amsmath,amssymb,graphicx,float,colordvi,url,wrapfig}
\usepackage[dvips]{color}
\usepackage{wrapfig}
\allowdisplaybreaks
\def\Tr{{\rm Tr}}

\begin{document}

%\thispagestyle{empty}
%\addtocounter{page}{-1}{} 
%\vskip-5cm

%\begin{flushright}
%KEK-TH-1558\\
%CAS-KITPC/ITP-326\\
%\end{flushright}

%\preprint{KEK-TH-XXXX}
%\preprint{CAS-KITPC/ITP-326}

\title{Hagedorn Instability in Dimensionally Reduced Large-$N$ Gauge Theories \\
as Gregory-Laflamme and Rayleigh-Plateau Instabilities}

\author{Takehiro {\sc Azuma}}
\email[]{azuma(at)mpg.setsunan.ac.jp}
\affiliation{
{\it Setsunan University,
17-8 Ikeda Nakamachi, Osaka, 572-8508, Japan }}

\author{Takeshi {\sc Morita}}
\email[]{tmorita(at)post.kek.jp}
%\affiliation{ {\it High Energy Accelerator Research Organization (KEK), Ibaraki 305-0801, \rm JAPAN }}
\affiliation{{\it Department of Physics and Astronomy,
University of Kentucky, Lexington, Kentucky 40506, USA}}

\author{Shingo {\sc Takeuchi}}
\email[]{shingo(at)nu.ac.th}
\affiliation{{\it Shanghai Jiao Tong University, Shanghai 200240, China}}
%\affiliation{{\it The Institute for Fundamental Study Naresuan University, Phitsanulok 65000, THAILAND}}
\altaffiliation{Present address: The Institute for Fundamental Study Naresuan University, Phitsanulok 65000, Thailand}

\begin{abstract}
%\enlargethispage{1000pt}
It is expected that the Gregory-Laflamme (GL) instability in the black string in gravity is related to the Rayleigh-Plateau instability in fluid mechanics.
Especially, the orders of the phase transitions associated with these instabilities depend on the number of the transverse space dimensions, and they are of first and second order below and above the critical dimension.
Through the gauge-gravity correspondence, the GL instability is conjectured to be thermodynamically related to the Hagedorn instability in large-$N$ gauge theories, and it leads to a prediction that the order of the confinement-deconfinement transition associated with the Hagedorn instability may depend on the transverse dimension.
We test this conjecture in the $D$-dimensional bosonic $D0$-brane model
% \cite{Banks:1996vh}
using numerical simulation and the $1/D$ expansion, and confirm the expected $D$ dependence.
%Our results open a new way of understanding the dynamics of the large-$N$ gauge theories in terms of both gravity and fluid mechanics.

\end{abstract}

\maketitle

%%%%%%%%%%%%%%%%%%%%%%%%%%%%%%%%%%%%%%%%%%%%%%%%%%%%%%%%%%%%%%%%%%

%%%%%%%%%%%%%%%%%%%%%%%%%%%%%%%%%%%%%%%%%%%%%%%%%%%%%%%
\paragraph*{Introduction.---}
%%%%%%%%%%%%%%%%%%%%%%%%%%%%%%%%%%%%%%%%%%%%%%%%%%%%%%%
Understanding the Hagedorn natures and the related confinement-deconfinement (CD) transition is one of the most important  problems in  gauge theory.
Recently, the gauge-gravity correspondence~\cite{Maldacena:1997re,Itzhaki:1998dd, Morita:2013wfa} suggested the relationship  between the Hagedorn instabilities in the large-$N$ gauge theories and the Gregory-Laflamme (GL) instabilities~\cite{Gregory:1994bj} in gravity~\cite{Martinec:1998ja, Aharony:2004ig, Kawahara:2007fn,  Mandal:2011ws}.
 Also, the GL instabilities are related to the Rayleigh-Plateau (RP) instabilities in fluid mechanics~\cite{Cardoso:2006ks, Miyamoto:2008rd, Caldarelli:2008mv, Lehner:2010pn}. Given these relations, the Hagedorn instabilities in the large-$N$ gauge theories are expected to have similarities to the GL and RP instabilities. 
The aim of this Letter is to shed light on this insight by studying the large-$N$ gauge theories.

In gravity, when we consider a background spacetime ${\mathbb R}^{D-1,1} \times S^1$, 
we obtain the uniform black string (UBS) solution whose event horizon winds on the $S^1$ space uniformly.
In this solution, if we increase the size of the $S^1$ space with the fixed mass, 
the horizon of the UBS is stretched and the GL instability arises above a critical size~\cite{Gregory:1994bj}. 
This instability makes the horizon of the black string nonuniform, and the GL transition occurs.
One significant property of the GL transition is that the order of the  transition depends on the number of the transverse dimension $D$ \cite{Sorkin:2004qq, Kudoh:2004hs, Kudoh:2005hf, Figueras:2012xj}. 
At $D \le 12$, we have a discontinuous first-order transition, and the stable solution at the critical size is the localized black hole (LBH) ($D\le 10$) or the nonuniform black string (NUBS) ($D=11,12$).
At $D \geq 13$, the second-order transition to the NUBS 
occurs.
Interestingly, if we fix the Hawking temperature rather than the mass, the order of the transition is first at $D \le 11$ and second at $D \ge 12$.
(See a review \cite{Kol:2004ww}.)

Remarkably, this instability is similar to the RP instability in fluid mechanics. 
Consider an extended fluid in ${\mathbb R}^{D-1,1} \times S^1$ with the same configuration as the event horizon of the UBS.
If we increase the size of the $S^1$ by fixing the volume, the RP instability arises above a critical size and  the fluid tends to be nonuniform.
The order of the phase transition associated with this instability depends on the dimension $D$.
It is of first order at $D \le 11$ and second order at $D \ge 12$~ \cite{Cardoso:2006ks, Miyamoto:2008rd}.
Thus, again, the transition becomes of higher order as $D$ increases.
Such a similarity between the fluid and the  black hole horizon may imply the existence of the particlelike black hole microstate~\cite{Morita:2013wfa} or the membrane paradigm~\cite{Thorne:1986iy}.

Since the gauge-gravity correspondence predicts that the GL transition is thermodynamically related to the CD transition of the large-$N$ gauge 
theories~\cite{Martinec:1998ja, Aharony:2004ig, Kawahara:2007fn,  Mandal:2011ws}, we expect a similar $D$ dependence there.
To test this conjecture, we study a $D0$-brane system from the gauge theory side
\cite{Aharony:2004ig, Kawahara:2007fn, Hanada:2007wn, Azuma:2007fj, Azeyanagi:2009zf, Mandal:2009vz, Catterall:2010fx, Azuma:2012uc}
 and investigate the $D$ dependence of the CD transition using Monte Carlo (MC) simulation for small $D$ and $1/D$ expansion for large $D$ \cite{Mandal:2009vz}.

\begin{figure*}
\vspace*{-8mm}
\begin{center}
\includegraphics[scale=1.0]{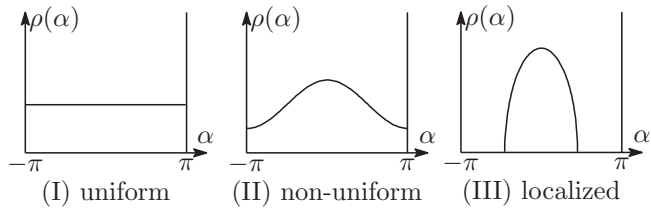}
\vspace*{-3mm}
\caption{The schematic plots of the eigenvalue distribution of $A_1$ defined by $\rho(\alpha) \equiv (1/N) \sum_{j=1}^N \delta(\alpha- L' \alpha_j) = (1/2\pi)( 1+ \sum_{n \neq 0} u_n e^{-in \alpha}) $ at large $N$.
}
\label{Fig-eigen-dens} 
\vspace*{-9mm}
\end{center}
\end{figure*}

\paragraph*{GL as CD transition.---}
First, we show how the GL and CD transitions are thermodynamically related in the gauge-gravity correspondence.
We see this connection in $N$ $D0$-branes in ${\mathbb R}^{8}\times S^1_\beta \times S_L^1$~\cite{Aharony:2004ig}.
Here, $S^1_\beta$ is the thermal temporal circle with the period $\beta$ and $S^1_L$ is the spatial circle with the period $L$.
We take $x^1$ as the spatial circle coordinate. 
This system at large $N$ in the strong coupling is described by classical supergravity,
which has the following two solutions: the smeared black $D0$-brane solution describing the uniformly aligned $D0$-branes  (UBS)
and the black hole solutions describing the $D0$-branes localized on the $S^1_L$ circle (LBH).
The UBS and LBH are stable for small and large $L$, respectively, and the first-order GL transition  occurs between them~\cite{Aharony:2004ig}.

This system is also described by the one-dimensional $SU(N)$ super Yang-Mills (1d SYM) theory~\cite{Taylor:1996ik}.
This theory involves the $N \times N$ adjoint scalars $X^I$ ($I=1,2,\cdots,9$) whose eigenvalues represent the positions of the $N$ $D0$-branes
in ${\mathbb R}^{8} \times S_L^1$.
The eigenvalues of $X^1$ are between 0 and $L$ because of the compactification. 
Then, the GL transition is interpreted by the transition from the uniform to the localized distribution of the $X^1$'s eigenvalues. (This is a large-$N$ phase transition and is smoothed out at finite $N$.)
This transition is identical to the CD transition through the T duality along the $x^1$ direction as follows.
This T duality maps the $D0$-brane model to the $D1$-brane model [the 2d $SU(N)$ SYM theory on the dual $S^1_{L'}$ whose periodicity is $L'=(2\pi)^2 \alpha'/L$, where $\alpha'$ is the Regge parameter]~\cite{Taylor:1996ik}
%\vspace*{-3.5mm}
\begin{align}
S  = 
\frac{N}{\lambda} \int_0^\beta \hspace{-2mm} dt  &  \int_0^{L'} \hspace{-2mm} dx \ 
\Tr 
\biggl\{ 
\frac12 F_{01}^2+ 
\sum_{I=2}^9 
\frac{1}{2} 
\left( 
D_\mu X^I \right)^2  \nonumber  \\
 &   - \sum_{I,J=2}^9 \frac{1}{4} [X^I,X^J]^2 
 +  \text{ fermions }  \biggr\}.   \label{2dSYM} 
\end{align}
%
%\vspace*{-3mm}

\noindent
Here, $\lambda$ is the 't Hooft coupling of the 2d SYM theory, $F_{01}$ is the field strength and $D_\mu$ is the covariant derivative.
 The adjoint scalar $X^1$ has been mapped to the gauge field $2 \pi \alpha' A_1$. 
 Thus, before and after the GL transition, the configuration of the eigenvalues of $A_1$ changes from the uniform distribution between  $0$ and $L/2\pi \alpha'(=  2\pi/L')$ to the localized one.
[(I) and (III) of Fig.~\ref{Fig-eigen-dens}.] 
If we take the static diagonal gauge $(A_1)_{ij} = \alpha_j \delta_{ij} $ ($i,j=1,\cdots, N$), the Polyakov loop along the $x^1$ circle, which is the order parameter of the deconfinement~\footnote{It may be more proper to call this transition the $Z_N$ breaking rather than the deconfinement.
However, we use the latter in this Letter, since they are equivalent in the Euclidian path integral.}, is written as $u_1 = 
 \sum_{j=1}^N  e^{i  L' \alpha_j}/N$.
Then, we easily see that $\langle |u_1| \rangle =0$ and $\langle |u_1| \rangle \neq 0$ in the uniform and localized distribution, respectively.
Thus, the GL transition can be interpreted as the CD transition, and the confinement and the deconfinement phases correspond to the UBS and LBH.
Indeed, this connection has been tested numerically in Ref. \cite{Catterall:2010fx}.

\paragraph*{Our Model.---}
We take the high-temperature limit of the 2d SYM theory (\ref{2dSYM}) \cite{Aharony:2004ig}. 
Then, the thermal Kaluza-Klein nonzero modes and fermions are classically decoupled and the theory reduces to the following model at $D=9$: 
%\vspace*{-4mm}
%\begin{align}
\begin{eqnarray}
\label{BFSS} 
\hspace*{-2mm} 
 S
=
\int_0^{L'} \hspace{-3mm} dx  
\Tr 
\Biggl\{ 
\sum_{I=1}^D
\frac{1}{2} 
\left( 
D_1 X^I \right)^2
-
\sum_{I,J=1}^D \frac{g^2}{4} [X^I,X^J]^2
\Biggr\}. 
%\end{align} 
\end{eqnarray}
%
%\vspace*{-4mm}
\noindent
Here, the coupling constant $g$ and adjoint scalars $X^I$ have been rescaled from (\ref{2dSYM}), and the gauge field $A_0$ has become one of the adjoint scalars $X^I$.
To investigate the $D$ dependence of the CD transition, we assign various values to $D$ in the model (\ref{BFSS}).
Since the adjoint scalars $X^I$ describe the $D0$-branes' positions, it is natural to compare the $D$ dependence of the GL transition in ${\mathbb R}^{D-1} \times S^1_\beta \times S_L^1$ and the CD transition of the model (\ref{BFSS}).

However, we cannot expect that the $D$ dependence of the GL transitions in Refs.~\cite{Sorkin:2004qq, Kudoh:2004hs, Kudoh:2005hf, Figueras:2012xj} and that of our model (\ref{BFSS}) agree exactly.
This is because Refs.~\cite{Sorkin:2004qq, Kudoh:2004hs, Kudoh:2005hf, Figueras:2012xj} studied
 the neutral black strings which are distinct from the smeared black $D0$-brane solution.
Besides, the model (\ref{BFSS}), except for $D=9$, is not related to the $D0$-branes in the superstring theory.
Thus, we cannot apply the gauge-gravity correspondence \cite{Itzhaki:1998dd}, and it is 
no wonder that no quantitative agreement is found.
However, we can still interpret our model as interacting $N$ particles in ${\mathbb R}^{D-1}  \times S^1_\beta \times S^1_L$ at high temperature and may show some fluid behaviors.
Hence, we study the model (\ref{BFSS}) focusing on the qualitative tendency of the $D$ dependence of the order of the phase transitions.

\begin{table*}
\vspace*{-5mm}
\begin{center} 
%{\scriptsize
\begin{tabular}{|c||c|c|c|c|c|c|}
\hline
$D$  & 2  & 3 & 9 & 15 & 20 & 2(with $b'|u_1|^4$) \\ \hline \hline
%$b'$ & 0  & 0 & 0 & 0  & 0 & 0.05 \\ \hline \hline
$1/L_c'$(MC) & 
1.3175  &
1.0975 &  
0.901 & 
0.884 & 
0.884 & 
1.3500 \\
\hline
$p$ & 
$1.05(3)$ &    % pHL4            = 1.04586          +/- 0.03252      (3.109%)
$1.00(1)$ &      % pHL6            = 0.999404         +/- 0.007067     (0.7071%)
$1.01(4)$ &    % pHL1            = 1.0073           +/- 0.03687      (3.661%)
$1.12(14)$ &  % pHL2            = 1.12442          +/- 0.1444       (12.85%)
$0.92(9)$ &
$0.81(5)$ % 0.810945    0.0543
 \\  % pHL3            = 0.917823         +/- 0.09227      (10.05%)
\hline \hline
$1/L_{H}^{'(1/D)}$ & 1.4(4) & 1.1(1) & 0.89(1) & 0.879(4) & 0.883(2)  & 1.4(4) \\
\hline
\end{tabular}
%}
\vspace*{-2mm}
\caption{
The critical point $L'_c$ and the exponent $p$ in the fitting (\ref{chi_fit}). 
%We take $b'=0$, except for the column ``$2(b'=0.05)$".
$L^{'(1/D)}_{H}$ is the Hagedorn point obtained by the $1/D$ expansion (\ref{tc1}), whose error is estimated as $\displaystyle 1/(L_H^{'(1/D)} D^2)$. The first-order transition is predicted to occur at $L'$ slightly larger than $L_{H}'$, and the closeness of $L'_c$ and $L^{'(1/D)}_{H}$ indicates that the $1/D$ expansion predicts the Hagedorn point well. }
\label{Tab-results}
\vspace*{-7mm}
\end{center}
\end{table*}

%\vspace*{-3mm}
\begin{figure*}
\vspace*{-2mm}
\begin{center} 
\includegraphics[scale=0.555]{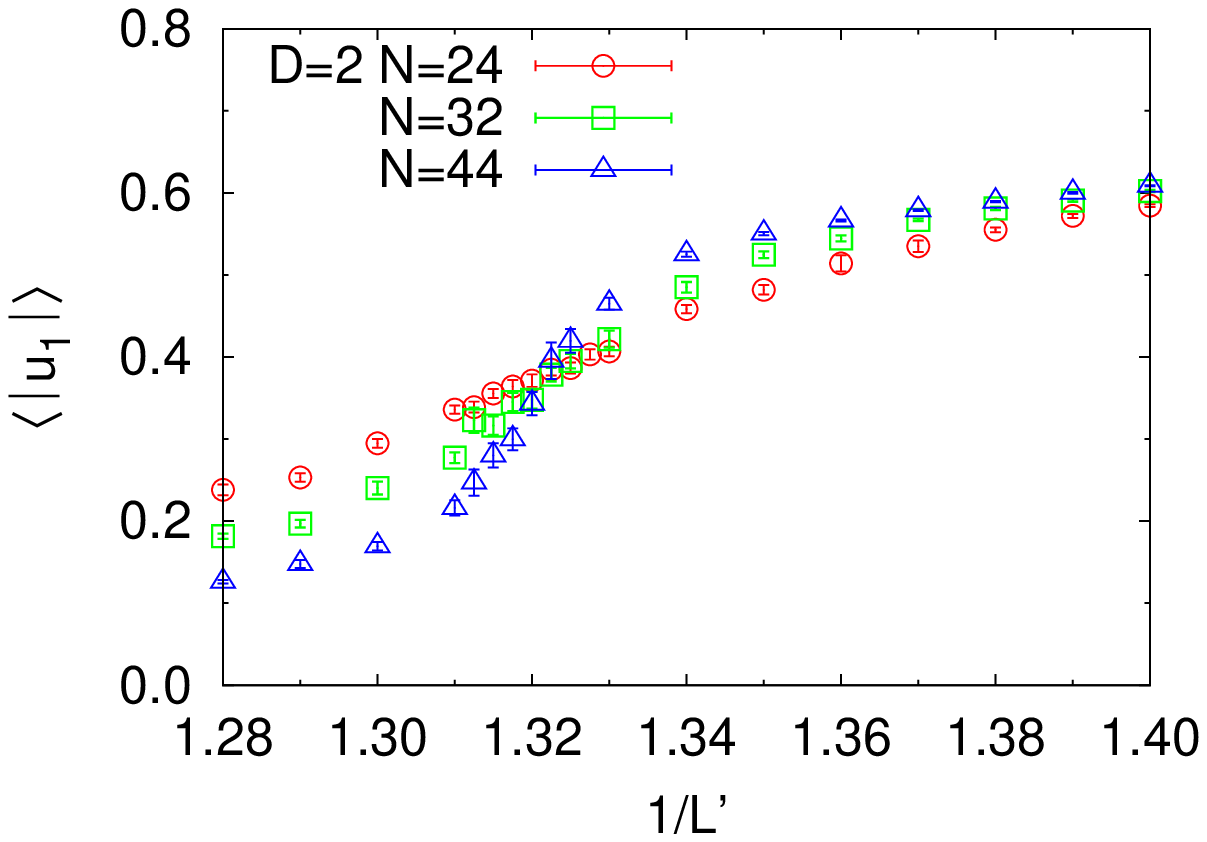}
\includegraphics[scale=0.555]{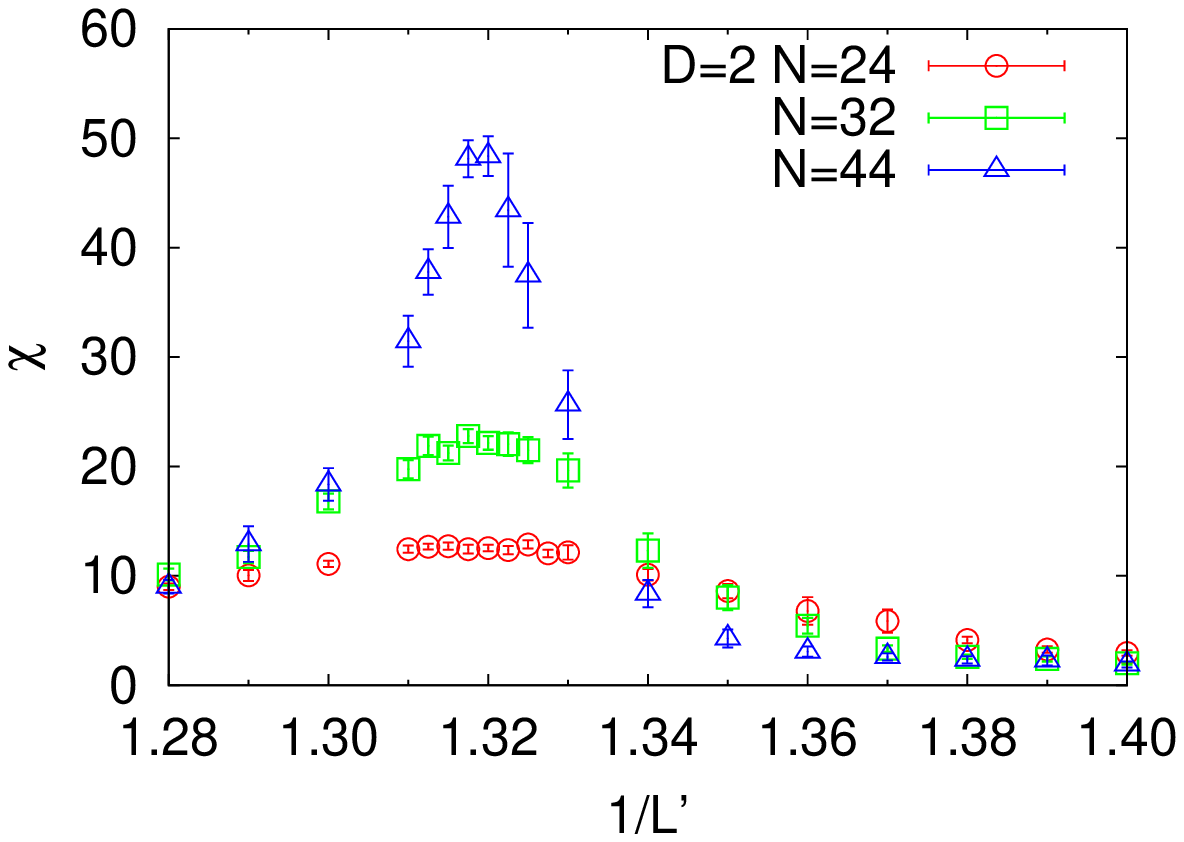}
%\vspace*{+1mm}
\vspace*{-3mm}
\caption{
The $L'$ dependence of $\langle |u_1| \rangle$ and $\chi$ at $D=2$.
We read off the critical point $L'_c$ at large $N$ from the peak of $\chi$.
%(Left-Bottom) $N^2$ dependence of $\chi$. We fit it as Eq.~(\ref{chi_fit}).
% and fix the exponent $p$ as listed in Table \ref{Tab-results}.
}
\label{fig-D2}
\end{center}
\vspace*{-8mm}
\end{figure*}
\paragraph*{Analysis through the $1/D$ Expansion.---}
The model (\ref{BFSS}) at large $D$ has been studied through the $1/D$ expansion \cite{Mandal:2009vz}, and we summarize the results.
At large $D$, we can integrate out the adjoint scalars $X^I$ \cite{Mandal:2009vz, Hotta:1998en} and obtain the effective action for the gauge fields $A_1$ as
%\vspace*{-1mm}
\begin{align}
\label{effective-action}
\textstyle S_{\text{eff}}= N^2 \Bigl( a_1 |u_1|^2+b|u_1|^4  + \sum_{n\ge 2} a_n |u_n|^2 + \cdots \Bigr) ,
\end{align}
%
%\vspace*{-2mm}
\noindent
where $u_n
\equiv 
 \sum_{j=1}^N  e^{i n L' \alpha_j}/N$ and we have taken the static diagonal gauge.
The coefficients $(a_n,b)$ are given by
%\begin{eqnarray}
%\textstyle \hspace*{-3mm}  
%\vspace*{-1mm}
\begin{align}
 a_1 =& 
 1- e^{-L' \tilde{\lambda}^{(1/3)}}
 \textstyle \left\{ D + L' \tilde{\lambda}^{(1/3)}  \left(
\frac{203}{160} -\frac{\sqrt{5}}{3} \right) \right\} \nonumber \\
& +\textstyle O\left(\frac{1}{D^2},\frac{1}{N^2}\right), \nonumber \\
 a_n =& {\textstyle \frac{1}{n}}- D e^{-nL' \tilde{\lambda}^{(1/3)}}   \textstyle + O\left(\frac{1}{D^{n}},\frac{1}{N^2}\right) \qquad \textrm{for } n \ge 2,
 \nonumber \\ 
%\textstyle \hspace*{-3mm}  
 b =& {\textstyle \frac{1}{3}} D L'\tilde{\lambda}^{(1/3)} 
e^{-2L' \tilde{\lambda}^{(1/3)}}
 \textstyle +  O\left(\frac{1}{D^2},\frac{1}{N^2}\right) ,
\label{b 1/D} 
\end{align} 
%
%
%\vspace*{-2mm}
%\end{eqnarray} 
\noindent 
where $\tilde{\lambda} =g^2 N D$, and we have taken $D \to \infty$, $N \to \infty$ and $g \to 0$ by fixing $\tilde{\lambda}$ finite.
The expressions (\ref{b 1/D}) are valid for large $L'$, so that $e^{-L' \tilde{\lambda}^{(1/3)}} \lesssim 1/D$.

Then, we can read off the phase structure of the model (\ref{BFSS}) from the effective action (\ref{effective-action}).
For large $L'$ (small $L$), since $a_n$ are positive for all $n$ , $|u_n|=0$ is stable.
There, the eigenvalue distribution of $A_1$ is uniform as depicted in (I) of Fig.~\ref{Fig-eigen-dens}  and the model is in the confinement phase, which corresponds to the UBS in gravity.
As we decrease $L'$, from (\ref{b 1/D}), the coefficient $a_1$ reaches 0 at 
%\vspace*{-2mm}
\begin{align} 
\textstyle
L_{H}'    = \frac{ \log D}{\tilde{\lambda}^{(1/3)}} 
\left\{ 1 +\frac{1}{D} \left( \frac{203}{160} -\frac{\sqrt{5}}{3}
\right) \right\} + O\left(\frac{\log D}{D^2} \right),
\label{tc1}
\end{align}
%
%\vspace*{-2mm} 
\noindent
and $|u_1|=0$ becomes unstable. This is the Hagedorn instability of the model (\ref{BFSS}) and it triggers the deconfinement transition.
We call $L'_H$ the Hagedorn point.

Generally, the order of the CD transition in the effective action (\ref{effective-action}) with arbitrary parameters $( a_n,b) $  is determined by the sign of $b$ at the Hagedorn point ($\left.a_1 \right|_{L'=L'_H} =0$, $a_{2,3,\dots}>0$)~\cite{Aharony:2003sx, AlvarezGaume:2005fv}.
If $b$ is positive, the second-order transition occurs at $L'_H$ and the
stable solution becomes $|u_1| = \sqrt{-a_1/2b}$, $|u_{2,3,\cdots}|=0$.
Then, the eigenvalue distribution becomes nonuniform as shown in (II) of Fig.~\ref{Fig-eigen-dens}, which corresponds to the NUBS in gravity.
%Since $|u_1|\neq 0$, this is a deconfinement phase.
On the other hand, if $b$ is negative, the first-order transition occurs at $L'_c$, which is slightly larger than $L'_H$.
The stable configuration is either the nonuniform phase [(II) of Fig.~\ref{Fig-eigen-dens}] or the localized phase [(III) of Fig.~\ref{Fig-eigen-dens}] depending on the details of the effective action.

In our case, $b$ in Eq.~(\ref{b 1/D}) is positive and the model (\ref{BFSS}) at large $D$ has the second-order transition.
Therefore, the CD transition bears resemblance to the transitions in gravity and the fluid model at large $D$.

%
%\begin{figure}[!!!!!!h]
\begin{figure*}
\vspace*{-12mm}
\begin{center} 
\includegraphics[scale=0.555]{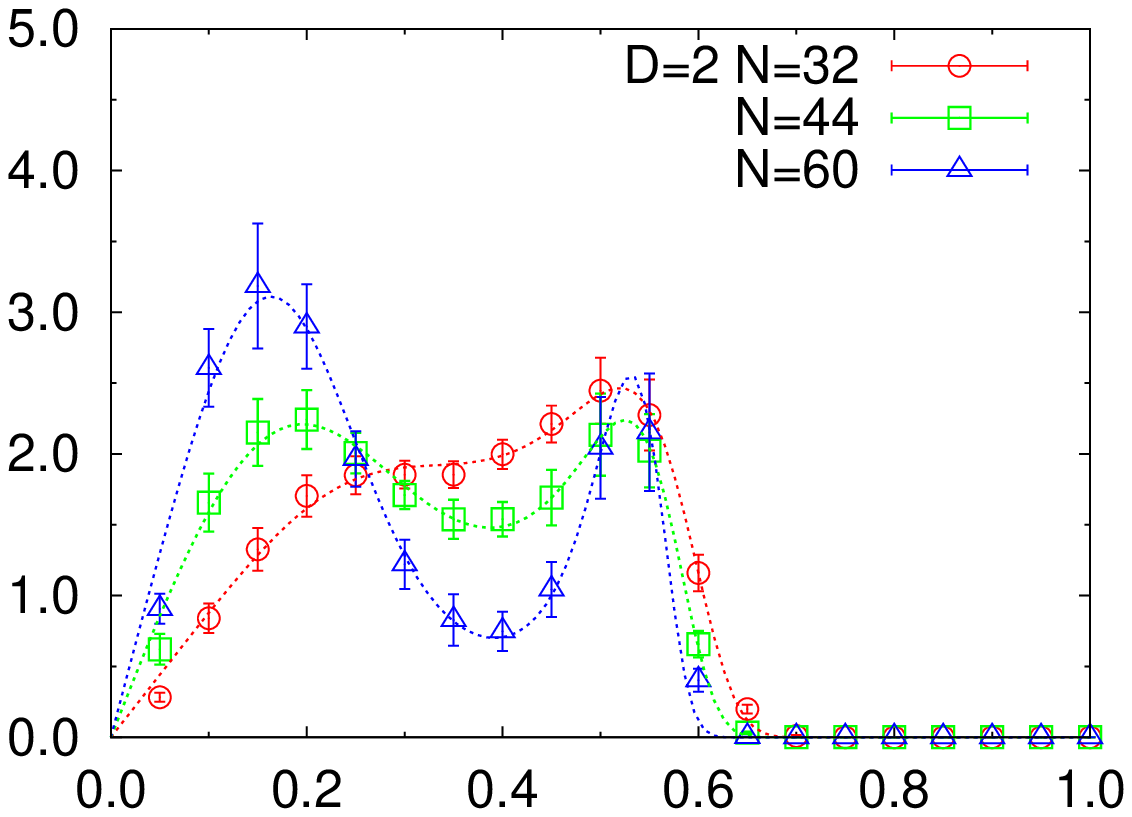}
\includegraphics[scale=0.555]{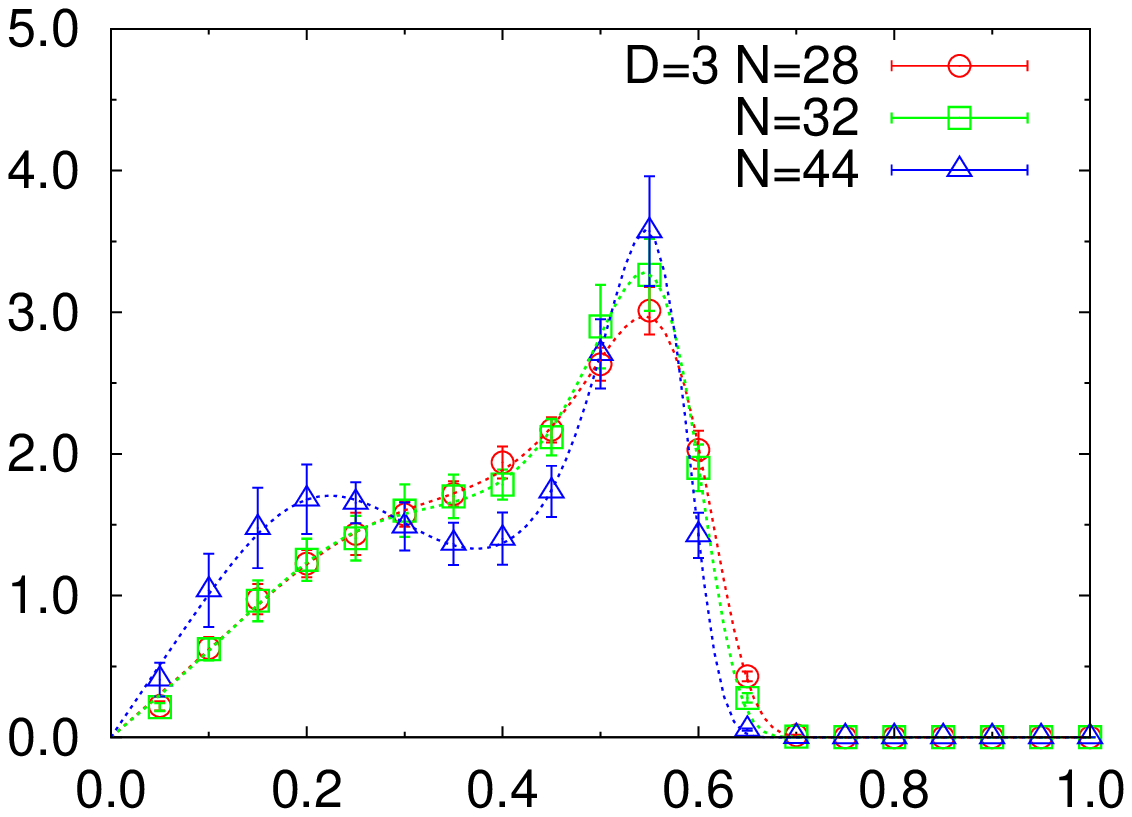}
\includegraphics[scale=0.555]{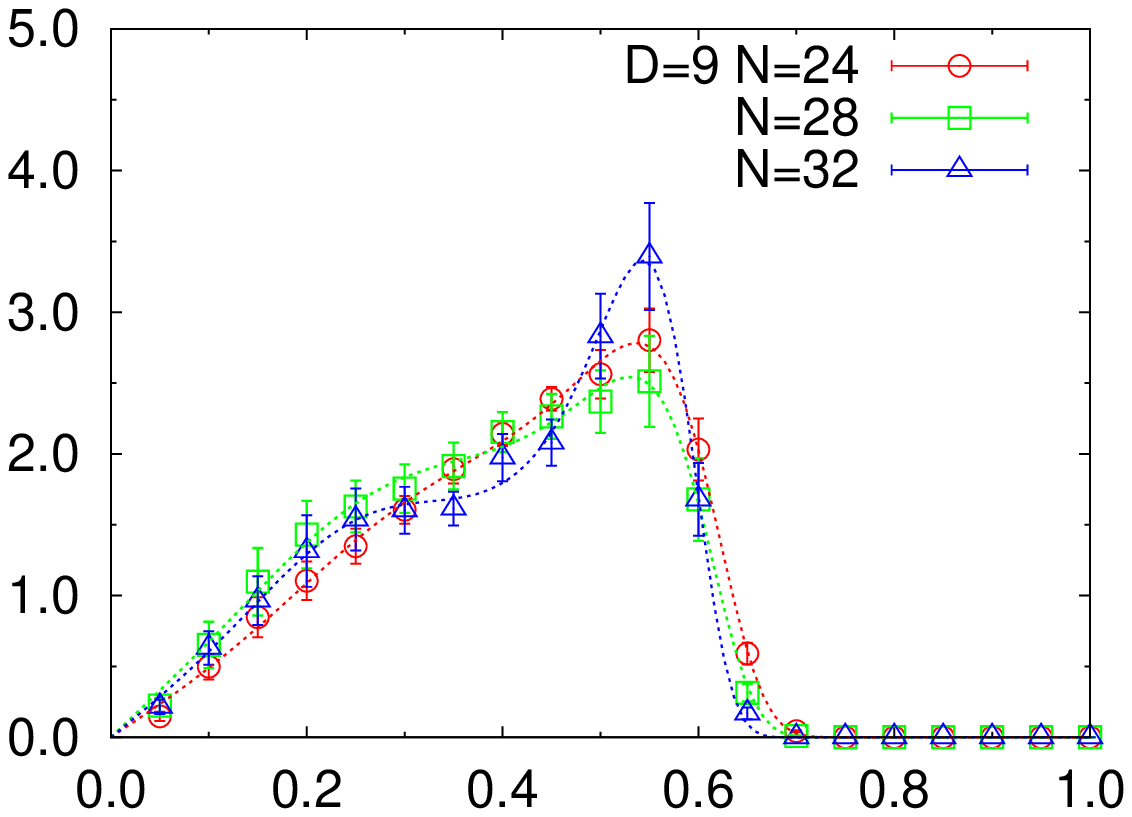}
\includegraphics[scale=0.555]{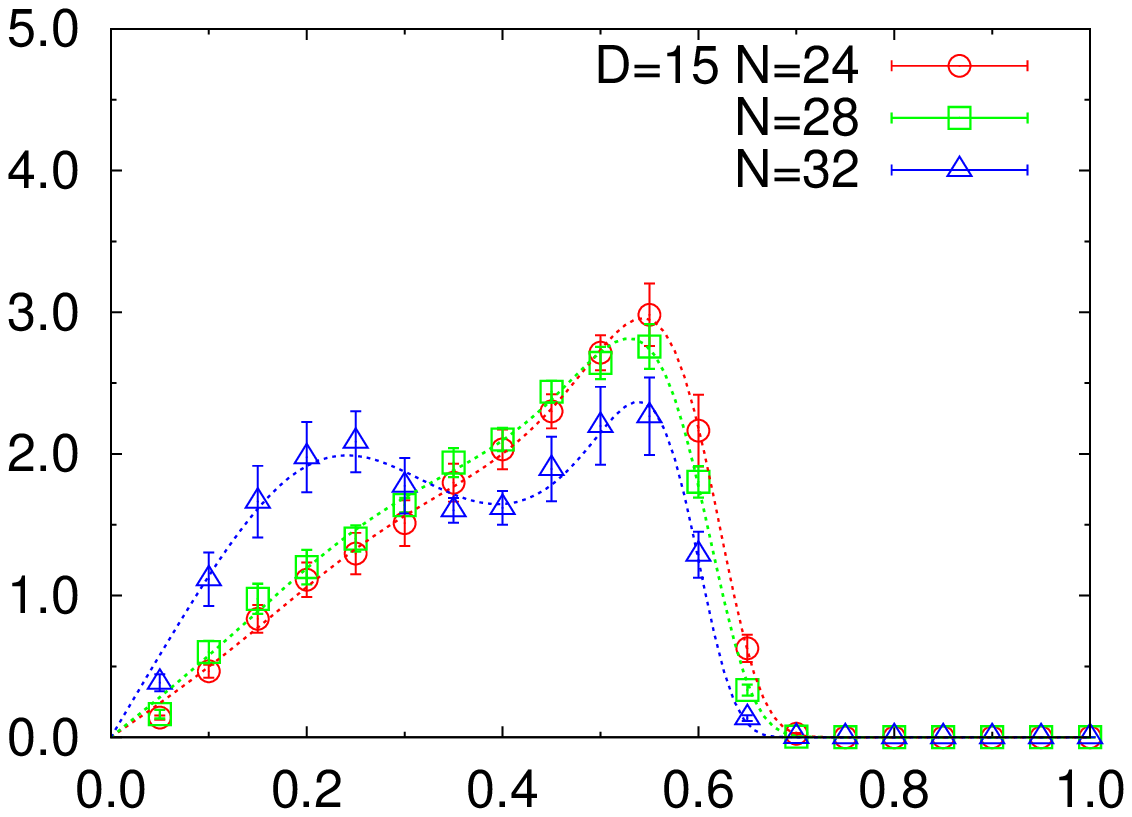}
\includegraphics[scale=0.555]{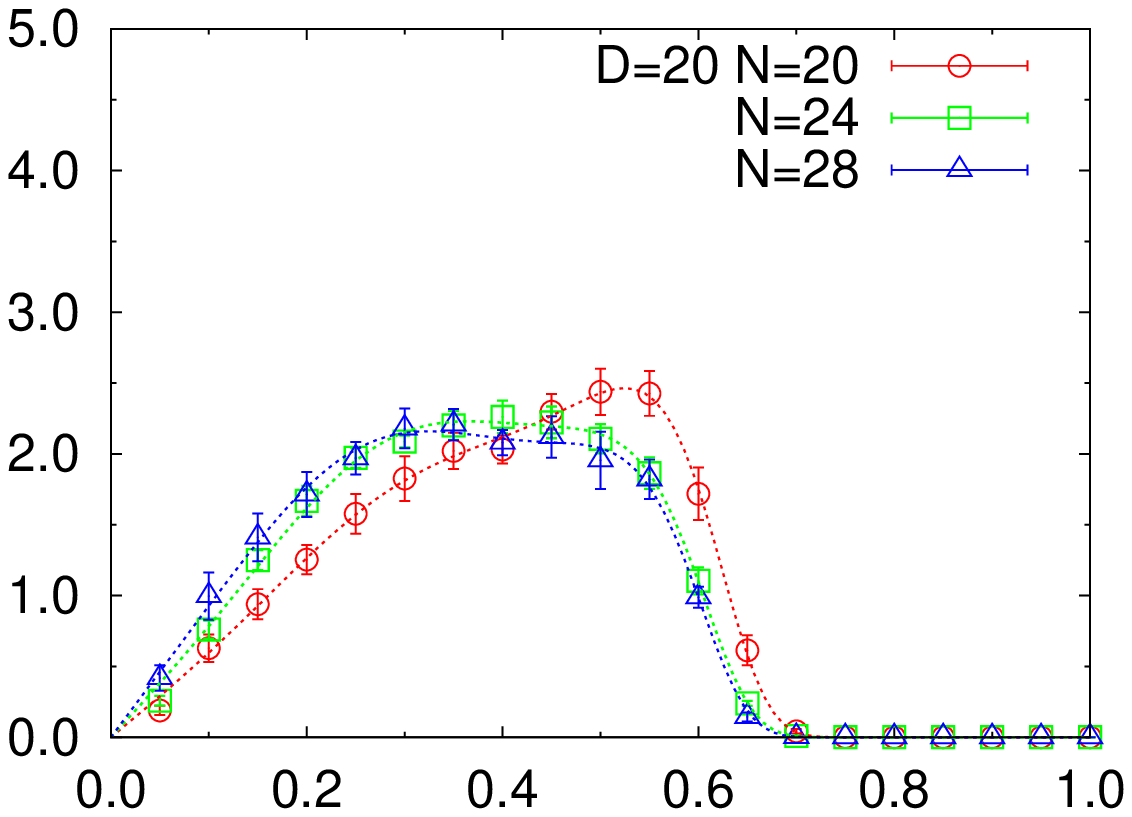}
\includegraphics[scale=0.555]{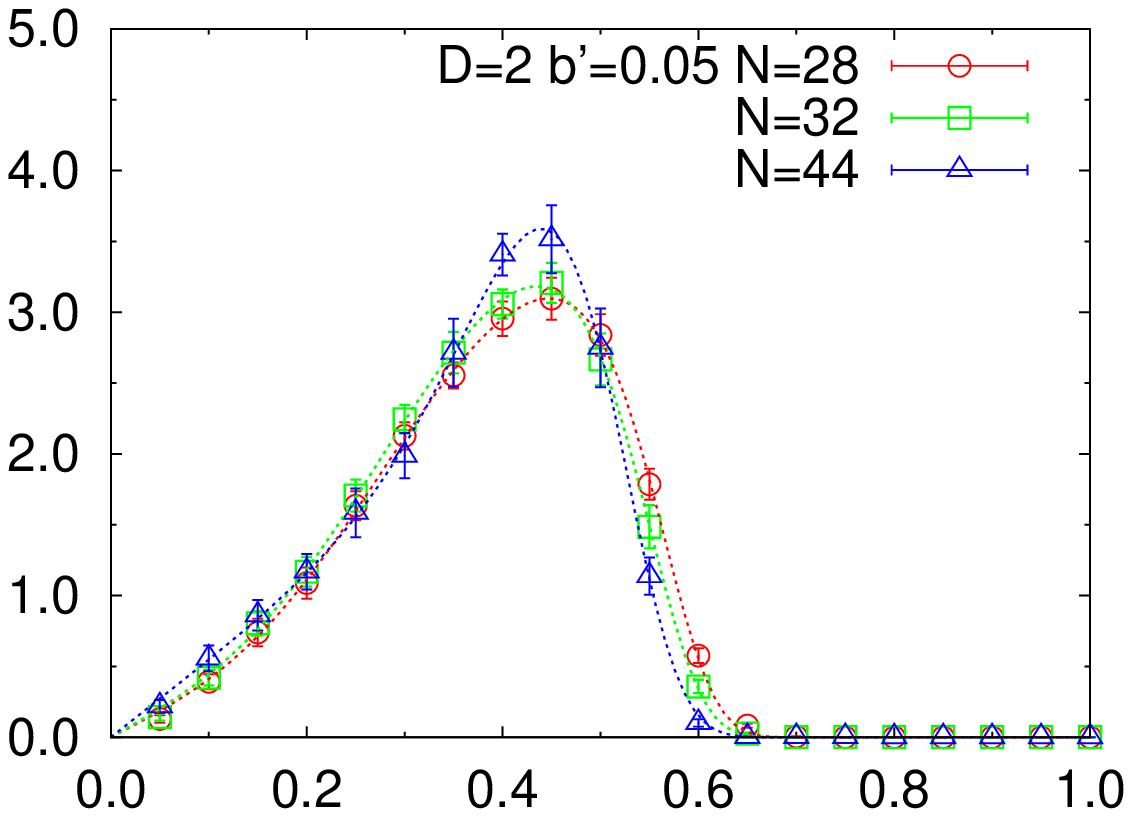}
\vspace*{-3mm}
\caption{
The density distribution of $|u_1|$ at the critical point for $D=2,3,9,15$, and $20$. 
The horizontal axis is $|u_1|$.
The right-bottom figure is the result of $D=2$ with $b'|u_1|^4$ ($b'=0.05$) for comparison, where the second-order transition occurs.
}
\label{figD02example}
\end{center} 
\vspace*{-7mm}
\end{figure*}
%\vspace*{-3mm}
\begin{figure*}
\vspace*{-2mm}
\begin{center} 
\includegraphics[scale=0.555]{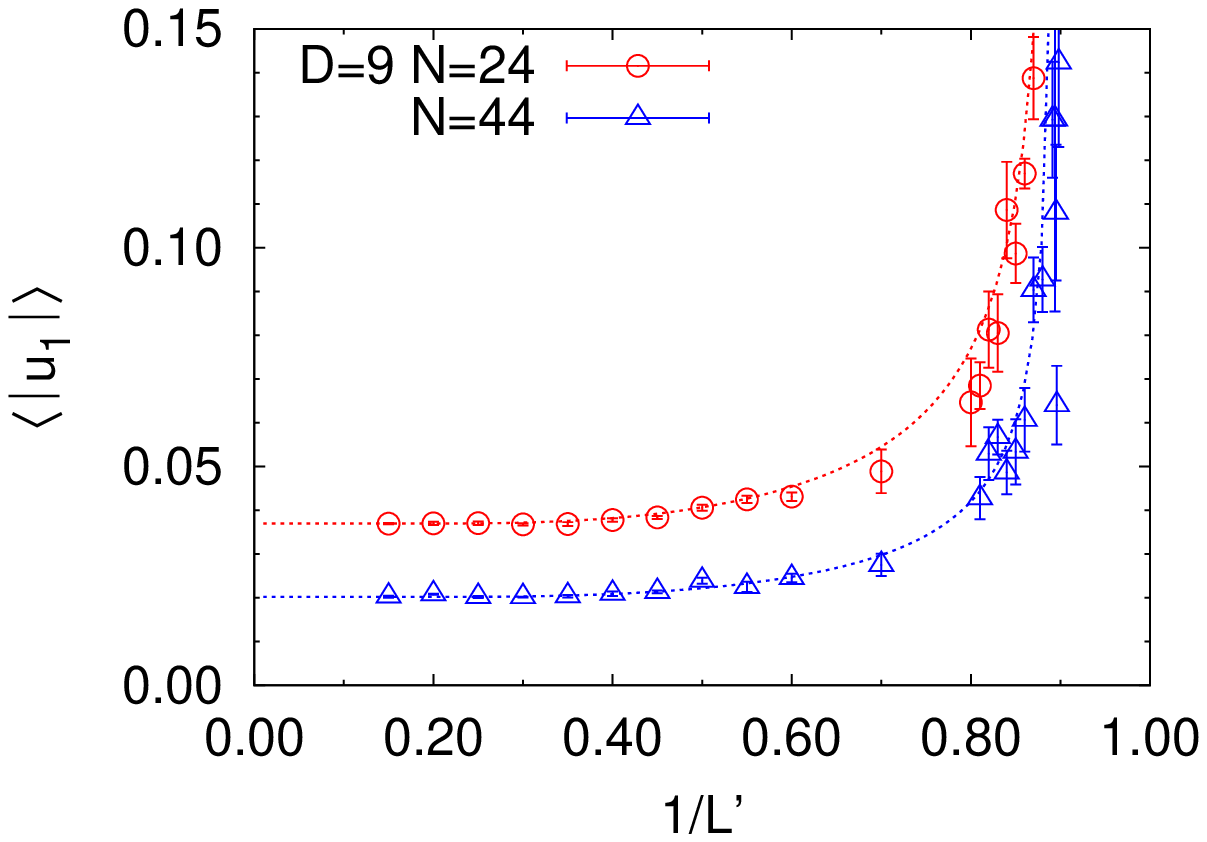}
\includegraphics[scale=0.555]{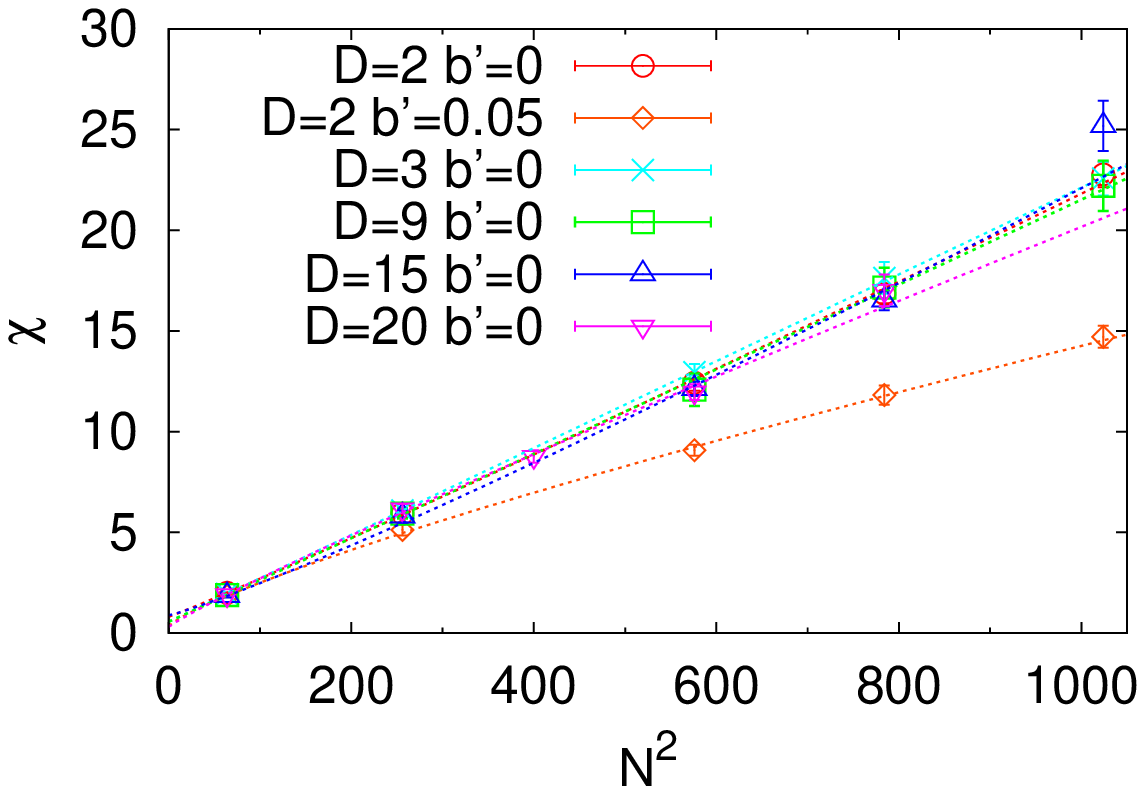}
%\vspace*{+1mm}
\vspace*{-3mm}
\caption{
(Left) The finite-$N$ effect of $\langle |u_1| \rangle$ in the confinement phase at $D=9$ ($1/L'_c=0.901$).
The curves are the prediction from the $1/D$ expansion (\ref{eq-u1-low-temp}).
(Right) The $N$ dependence of $\chi$ at the critical point fitted as Eq.~(\ref{chi_fit}).}
\label{Fig_chimax}
\end{center}
\vspace*{-8mm}
\end{figure*}

\paragraph*{Numerical Results for small $D$.---}
To investigate the model (\ref{BFSS}) for small $D$,
we perform the  hybrid MC lattice calculation at $D=2,3,9,15$, and $20$~\footnote{
References \cite{Aharony:2004ig,  Kawahara:2007fn, Azeyanagi:2009zf} studied  the phase transition of the model (\ref{BFSS}) numerically.
Reference \cite{Aharony:2004ig} figured out that $b$ in the effective action (\ref{effective-action}) is close to 0 at $D=9$, and the model is near the border of the first and second-order transition.
References \cite{Kawahara:2007fn} and \cite{Azeyanagi:2009zf} argued that $D=9$ and $D=2,3$ have higher-order transitions, respectively, by fitting  the observables as the functions of $L'$.
%with  $L'$.
However, such a fitting is difficult due to the large finite-$N$ effects near the critical point as argued in Eq.~(\ref{eq-u1-low-temp}), and we consider different approaches.
}.
We use a unit $g^2N=1$ and take the number of the lattice sites to be 15.
To specify the phase transition point at large $N$, we evaluate the Polyakov loop $|u_1|$ and its susceptibility
$\chi \equiv N^2\left(\langle |u_1|^2 \rangle-\langle |u_1| \rangle^2\right)$
%$ \chi = N^2\langle (|u_1|-\langle |u_1| \rangle )^2 \rangle$
 as plotted  in Fig.~\ref{fig-D2}.
Both the first and second-order transitions are expected to occur at the critical point $L'_c$ where $\chi$ takes the maximum. 
The obtained critical points are summarized in Table~\ref{Tab-results}. 

To evaluate the order of the  transition, we plot the density distribution of $|u_1|$ at the critical point in Fig. \ref{figD02example}. At $D=2,3,9$, and 15, we observe two peaks as $N$ grows larger.
The peak which lies at smaller and larger $|u_1|$ corresponds to the confinement and deconfinement, respectively.
This shows the existence of the metastable state, which is strong evidence for the first-order transition.
At $D=20$, due to the CPU cost, we have calculated only up to $N=28$ and
do not observe two peaks clearly.
However, the density distribution seems to consist of two wide peaks.
To ensure this point, we add the term $b'|u_1|^4$ with $b'=0.05$ to the action (\ref{BFSS}) at $D=2$ so that the model has the second-order transition. 
[Recall that the analysis of the effective action (\ref{effective-action}) shows that the transition tends to be of second order at $b'>0$.]
The obtained $|u_1|$ density distribution at the second-order critical point in Fig. \ref{figD02example} is
 a single wide peak, in contrast with the case of $D=20$.
Thus, we presume that the transition is still of first order at $D=20$
\footnote{
In the case of the first-order phase transition, 
there are two possible stable deconfinement phases: the nonuniform  and the localized phases.
In the density distribution of $|u_1|$ in Fig.~\ref{figD02example}, the peaks of the deconfinement phases are larger than $1/2$.
These might be the evidence that the deconfinement phases are the localized ones~\cite{Aharony:2004ig, Aharony:2003sx, AlvarezGaume:2005fv, Mandal:2009vz}. However, we have to extrapolate the results at large $N$ carefully, and 
we leave this issue for a future work.
}.

Note that the peaks of $|u_1|$ in Fig.~\ref{figD02example} in the confinement phase do not lie exactly at $|u_1|=0$, due to the finite-$N$ effect.
Using the $1/D$ expansion, we calculate the leading $1/N$ correction at large $L'$ ($ L' \gg L'_H $) from Eq.~(\ref{effective-action}) as
%\vspace*{-3mm}
\begin{align}
%\langle |u_1| \rangle = \frac{1}{Z} \int \prod_n du_n du_n^\dagger |u_1| e^{-S_{\text{eff}}}= \frac{1}{2N} \sqrt{\frac{\pi}{a_1}}
\langle |u_1| \rangle = \frac{\int \prod_n du_n du_n^\dagger |u_1| e^{-S_{\text{eff}}}}{\int \prod_n du_n du_n^\dagger e^{-S_{\text{eff}}}} \simeq \frac{1}{2N} \sqrt{\frac{\pi}{a_1}}.
\label{eq-u1-low-temp}
\end{align}
%
%\vspace*{-3mm}
\noindent 
Here, each $u_n$ can be treated as an independent variable and $b |u_1|^4$ can be neglected at large $N$ in the confinement phase \cite{Aharony:2003sx}.
This effect is significant near the Hagedorn point where $a_1$ is close to 0.
This result quantitatively agrees with the MC results as shown in Fig.~\ref{Fig_chimax} (Left).
% At $D=2$, $N=60$, $1/L' = 1.3175$, the peak of $\langle |u_1| \rangle$ in the confinement phase lies around $0.15$ in Fig. \ref{figD02example}, which is consistent with $ \frac{\sqrt{\pi/a_1}}{2N} \simeq 0.0747\cdots$.

As further evidence of the phase transition order, we fit the $N$ dependence of $\chi$ at the critical point as 
%\vspace*{-3mm}
\begin{eqnarray}
\left.\chi\right|_{L'=L'_{c}} = \gamma_1 N^{2p} + \gamma_2, \label{chi_fit}
\end{eqnarray}
%
%\vspace*{-3mm}
\noindent 
with the fitting parameters $(\gamma_1, \gamma_2, p)$. The analysis of the effective action (\ref{effective-action}) at large $N$ shows that the exponent $p$ is $1$ and $1/2$ for the first ($b<0$) and second order ($b>0$), respectively. Thus, we can distinguish the transition order by $p$~\cite{Fukugita:1990vu}.
The result is summarized in Table \ref{Tab-results} and Fig.~\ref{Fig_chimax} (Right). 
For $D \le 15$, $p$ is close to 1, which is consistent with the first-order transition.
At $D=20$, we have $p= 0.92(9)$, which is not decisive due to error.
However, it differs from the second-order case ($D=2$ with $b'|u_1|^4$), where $p=0.81(5)$, and would be  consistent with the first-order transition.

Therefore, we conclude that the CD transitions in the model (\ref{BFSS}) are of first order until at least $D=15$ (presumably until $D=20$), which is again consistent with gravity and the fluid model which have the first-order transition at small $D$
~\footnote{
The MC calculation shows that the transition orders until $D=15$ (or 20) are different from the $1/D$ expansion.
However, the Hagedorn point $L'_H$  in Table \ref{Tab-results} and $\langle |u_1| \rangle$ in the confinement phase in Fig.~\ref{Fig_chimax} (Left) quantitatively agree.
Further, it has been investigated that the $1/D$ expansion works at small $L'$ ($L' \ll L'_H$) \cite{Azuma:2012uc,Hotta:1998en}.
These results may imply that the $1/D$ expansion properly predicts $a_n$ in the effective action (\ref{effective-action}), but $b$ is different.
Equation (\ref{b 1/D}) shows that $b$ near the Hagedorn point in the $1/D$ expansion is an $O(1/D)$ quantity. 
The numerical calculation also shows that $b$ is small~\cite{Aharony:2004ig}.
Thus, $b$ might be sensitive to some nonperturbative effects in the $1/D$ expansion.
In that case, when $D$ is so small that such nonperturbative effects are significant, the CD transition in the $1/D$ expansion might be of first order.}.

\paragraph*{Conclusions.---}
We have studied the $D$ dependence of the CD transition in the model (\ref{BFSS}) and observed that it is of first and second order at small and large $D$.
This tendency of the $D$ dependence is similar to  the transitions in gravity  and the fluid model~\cite{Cardoso:2006ks, Miyamoto:2008rd} and is strong evidence that the Hagedorn instability in the model (\ref{BFSS}) is related to the GL and RP instabilities.
It is important to study larger $D$ numerically until we reach the critical dimension where the transition switches from the first to the second order to confirm this relation.

One reason why our model shows the fluid natures may be that it describes the $N$ $D0$-branes in ${\mathbb R}^{D-1}\times S^1_\beta  \times S^1_{L}$ which may compose the fluid.
However, $D$ is merely the dimension of the internal $SO(D)$ symmetry (flavor) of the adjoint scalars $X^I$ and is not a specific parameter of our model.
Besides, the dual spatial circle $S^1_{L'}$ in the model (\ref{BFSS}) can be regarded as the thermal temporal circle.
Therefore, some fluid interpretations may be widely applied to other finite-temperature large-$N$ gauge theories, too.
It will open a new possibility that not only the gravities but also the fluid models illuminate the dynamics of the large-$N$ gauge theories.

%Lastly, we comment on the validity of the $1/D$ expansion.
%The MC calculation shows that the order of the transition differs from that obtained by the $1/D$ expansion until $D=15$ (or 20).
%However, the Hagedorn point $L'_H$  in Table \ref{Tab-results} and the $\langle |u_1| \rangle$ in the confinement phase in Fig.~\ref{Fig_chimax} (Left) quantitatively agree.
%Further, it has been investigated that the $1/D$ expansion works at small $L'$ ($L' \ll L'_H$, the deconfinement phase) \cite{Azuma:2012uc,Hotta:1998en}.
%These results may imply that the $1/D$ expansion properly predicts $a_n$ in the effective action (\ref{effective-action}) but $b$ is different.
%Equation (\ref{b 1/D}) shows that $b$ near the Hagedorn point in the $1/D$ expansion is an $O(1/D)$ quantity. 
%Numerical calculation also indicates that $b$ is small \cite{Aharony:2004ig}.
%Therefore, $b$ might be sensitive to some nonperturbative effects in the $1/D$ expansion.
%In that case, when $D$ is so small that such nonperturbative effects are significant, the CD transition in the $1/D$ expansion may become of first order.

We thank S.~Hashimoto, E.~Itou, G.~Mandal, and J.~Nishimura for valuable discussions and comments, and K.~N.~Anagnostopoulos and S.~Nishida for technical support.
The numerical simulations were performed at KEKCC and NTUA het clusters. 
We are supported in part by JSPS (Grants No. 23740211 for T.A. and No. 24840046 for T.M.) and NFS (Grant No. NSF-PHY-1214341 for T.M.).

%The authors thank Shoji Hashimoto, Etsuko Itou, Gautam Mandal, Keiju Murata, Jun Nishimura, Muneto Nitta, Shinji Tsujikawa and the members of the Particle Physics Group in Tokyo Institute of Technology for valuable discussions, and Konstantinos N.~Anagnostopoulos and Shohei Nishida for technical supports. The numerical simulations were performed at KEKCC and NTUA het clusters. The work of T.A. and T.M. is supported in part by JSPS (Nos. 23740211 and 24840046 resp.).

%Grant-in-Aid for Scientific Research 

\end{document}